\documentclass[aps,floats,showpacs,showkeys]{revtex4}
\usepackage{amssymb}
\usepackage{amscd}
\usepackage{amsfonts}
\usepackage{mathrsfs}
\usepackage{epsfig}

\usepackage{amsmath,amssymb}
\usepackage{graphicx}
\usepackage{verbatim}

\input vatola.sty 

\newcommand{\bastar}{\begin{eqnarray*}}
\newcommand{\eastar}{\end{eqnarray*}}

\newskip\humongous \humongous=0pt plus 1000pt minus 1000pt
\relax
\newcommand{\be}{\begin{equation}}
\newcommand{\ee}{\end{equation}}
\newcommand{\bea}{\begin{eqnarray}}
\newcommand{\eea}{\end{eqnarray}}
\newcommand{\pro}{\partial}
\newcommand{\ba}{\begin{array}}
\newcommand{\ea}{\end{array}}

\newcommand{\nn}{\nonumber}
\newcommand{\tF}{{\tilde F}}
\newcommand{\s}{\sum_{n=1}^{\infty}}

\begin{document}
\title{Birefringence and non-transversality of light propagation in an ultra-strongly magnetized vacuum}
\bigskip
\author{Shuang-Wei Hu}
\email{hu@mail.nankai.edu.cn} \affiliation{Theoretical Physics
Division, Chern Institute of Mathematics, Nankai University, Tianjin
300071, P.R.China}
\author{Bin-Bin Liu}
\affiliation{Theoretical Physics Division, Chern Institute of
Mathematics, Nankai University, Tianjin 300071, P.R.China}
\begin{abstract}
The birefringence phenomenon in the vacuum with a constant magnetic
background of arbitrary strength is considered within the framework
of the effective action approach. A new feature of the birefringence
in a magnetized vacuum is that the parallel mode, which is polarized
parallel to the plane containing the magnetic field and the photon
wave vector, is no longer transverse. We have studied this feature
in detail for arbitrary magnetic field and provided analytic results
for the ultra-strong magnetic field regime. Possible physical
implications of our results in astrophysics are discussed.

\end{abstract}
\vspace{0.3cm}  \maketitle
\bigskip

\section{Introduction}

The theoretical investigation of nonlinear effects on light
propagation, including vacuum birefringence, has been extensively
studied since early 70s \cite{birula,brezin,adler,shabad}.
Recent years have witnessed a significant growth of interest in this
realm of research \cite{daniels,dittgies,dittbook,salim,pak,heinzl},
especially in the vacuum birefringence in ultra-strong fields, due
to predictions of presence of strong magnetic fields in
astrophysical objects \cite{pulsar,astro,shaviv} and the
technological improvement in high-intensity laser fields
\cite{laser} above the critical strength $B_c=\dfrac{m^2c^2
}{e\hbar}\simeq 4.4\times10^{13}G$.
The birefringence phenomenon in magnetized media reveals a new
interesting feature related with the fact that the polarization
vector of the parallel mode of the propagating photon becomes
non-transverse, i.e., it fails to be orthogonal to the wave vector
\cite{brezin,dittbook}.
One way of investigating the vacuum birefringence is to work within
the effective Lagrangian approach. Recently the analytic series
representation for the one-loop effective action of quantum
electrodynamics (QED) has been obtained \cite{qed1} on the base of
Schwinger's integral expression for the effective action
\cite{schw}. This explicit analytical expression is helpful to
investigate the light propagation in magnetic field of arbitrary
strength, especially in strong magnetic fields of magnitude $B$
above the critical value $B_{c}$.

In the present paper we consider the birefringence in arbitrary
homogeneous magnetic field as well as the effect of light
non-transversality between the polarization
vector and the wave vector. Since this effect is small (it is of second order in
the fine structure constant) it was neglected for the field $B$
satisfying $0\leq B\leq {\cal O}(B_c)$ in previous studies
\cite{brezin,dittbook}. For ultra-strong magnetic field regime,
$B>>B_{c}$, one should expect this effect will affect the
propagation of light significantly. The purpose of our paper is to
investigate this effect in a detail for magnetic field of arbitrary
strength.

\section{Effective action formalism} \label{sec:general}

The effective action provides us with a useful bridge between the
full quantum theory and classical field theory. Once the effective
action is known, in soft photon approximation (photon energy smaller
than the electron mass), classical equations of motion can be
derived to describe the light propagation in the language of
classical physics.

Let us start with the main lines of the effective action approach to
light propagation in various vacua \cite{dittgies,visser}. An
integral expression for the one-loop effective action is given by
Schwinger \cite{schw}
\bea  {\cal L}_{\mathrm{eff}} &=& -x - \dfrac{1}{8\pi^2}
\int_{0}^{\infty} \dfrac{dt}{t^3} e^{-m^2 t} \Bigg{[} -
\dfrac{2}{3}(et)^2 x -1 \nn \\
&&+(et)^2 |y| \coth \Bigg{(} et\sqrt {\sqrt {x^2 + y^2} + x}
\Bigg{)} \cot \Bigg{(} et\sqrt {\sqrt {x^2 + y^2} - x} \Bigg{)}
\Bigg{]} , \label{eq:schw} \eea
where we have introduced the gauge and Lorentz invariants of the
electromagnetic field,
\bea && x = \dfrac{1}{4} F_{\mu \nu} F^{\mu \nu},\nn \\
&& y = \dfrac{1}{4} F_{\mu \nu} {\tilde F}^{\mu \nu},~~\tF^{\mu\nu}
= \dfrac{1}{2} \epsilon^{\mu \nu \rho\sigma} F_{\rho \sigma}. \eea
We employ $\epsilon^{0123}=-1$ and $\eta_{\mu
\nu}=\mathrm{diag}(-1,1,1,1)$. We use Greek letters for the
space-time indices $(0,1,2,3)$ and Latin letters for the spatial
ones. For convenience, we use natural units $\hbar=c=1$
throughout the paper.

To obtain exact analytic results we will use an exact series
representation for the one-loop effective Lagrangian of QED
\cite{qed1}
\bea {\cal L}&=& -\dfrac{a^2 - b^2}{2} \nn \\ &&- \dfrac{e^2
}{4\pi^3} ab \sum_{n=1}^{\infty}\dfrac{1}{n} \Big [\coth(\dfrac{n
\pi b}{a}) \Big ( {\rm ci}(\dfrac{n \pi m^2}{ea}) \cos(\dfrac{n \pi
m^2}{ea}) +{\rm si}(\dfrac{n \pi m^2}{ea}) \sin(\dfrac{n \pi
m^2}{ea})\Big ) \nn
\\ && -
\dfrac{1}{2} \coth ( \dfrac{n \pi a}{b}) \Big ( \exp(\dfrac{n \pi
m^2}{eb}) {\rm Ei}(-\dfrac{n \pi m^2}{eb} ) + \exp(-\dfrac{n \pi
m^2}{eb} ) {\rm Ei}(\dfrac{n \pi m^2}{eb} ) \Big )\Big ] ,
\label{L0} \eea
where $a,b$ are gauge invariant variables corresponding to the
magnetic and electric fields in an appropriate Lorentz frame,
\bea &&a = \sqrt {\sqrt {x^2 + y^2} + x}, \nn \\
&&b = \sqrt {\sqrt {x^2 + y^2} - x}. \eea

In the weak field limit the expansion of the integral in
(\ref{eq:schw}) produces the well-known Euler-Heisenberg effective
Lagrangian \cite{ritus,dittgies}
\bea &&\tilde {\cal L}_{\mathrm{E-H}} = -x+ \tilde{c}_1 x^2+\tilde{c}_2 y^2, \nn \\
&&  \tilde{c}_1 = \dfrac{8 \alpha^2}{45 m^4},
~~~~\tilde{c}_2=\dfrac{14 \alpha^2}{45 m^4} , \label{eq:EH2} \eea
where $\alpha = \dfrac{e^2}{4 \pi \epsilon_0}=\dfrac{1}{137.036}$ is the fine
structure constant. In our choice of natural units we set
$\epsilon_0=1$, so that the corresponding value of
electron charge is $e=\sqrt {4\pi \alpha}$.

We will follow the effective action approach \cite{visser} to study
the light propagation effects in nonlinear electrodynamics. We
assume that the soft photon approximation, the linearization
procedure, and the restricted eikonal approximation make sense
\cite{visser}. It is suitable to split the total electromagnetic
field into the background field $F_{\mu \nu}$ and the propagating
photon $f_{\mu \nu}$ with the vector potential $a_{\mu}(k)$ and wave
vector $k^{\mu}$. We keep the linear approximation with respect to
$f_{\mu \nu}$ in equations of motion. After these two procedures the
equations of motion corresponding to the full effective action lead
to an eigenvalue equation for the propagating modes,
 \bea
 A^{\mu \nu}\epsilon_\nu =0 , \label{eigen1}
\eea
where $\epsilon_\nu \equiv a_{\nu}/(a_{\mu}a^{\mu})^{1/2}$ is a unit
polarization vector of the soft photon, the symmetric tensor $A^{\mu
\nu}$ is given by
 \bea A^{\mu
\nu} &\equiv& 2\left.{{\partial^2 {\cal L} \over\partial
F_{\mu\alpha} \;
\partial F_{\nu\beta}} }\right|_{\mathrm{background}}  k_{\alpha}
k_{\beta} \nn \\
&=& c_1 F^{\mu \alpha} F^{\nu \beta} k_{\alpha}k_{\beta}+c_2 \tilde
F^{\mu \alpha} \tilde F^{\nu \beta} k_{\alpha}k_{\beta}
+c_3(\delta^{\mu \nu} k^2-k^{\mu}k^{\nu})\nn \\
 && + c_5(F^{\mu \alpha}
 \tilde F^{\nu \beta}+\tilde F^{\mu \alpha}
F^{\nu \beta} )k_{\alpha}k_{\beta}, \eea
and the derivative functions are defined by
\bea c_1 \equiv \frac{1}{2} \partial_x^2 {\cal L} ,~~ c_2 \equiv
\frac{1}{2} \partial_y^2 {\cal L} ,~~c_3 \equiv
\frac{1}{2}\partial_x {\cal L},~~ c_4 \equiv \frac{1}{2}\partial_y
{\cal L}, ~~c_5 \equiv \frac{1}{2}\partial_{x y} {\cal L}.
\label{deri} \eea
It can be shown \cite{salim} that the equation (\ref{eigen1}) is
indeed equivalent to the light cone condition obtained in
\cite{dittgies} without using the averaging over polarization modes.

Solutions of equation (\ref{eigen1}) represent the dynamically
allowed polarization modes. Nontrivial solutions to this equation
exist if a generalized Fresnel equation is satisfied \cite{kremer}
 \bea \det
A^{\mu \nu}(k) = 0. \label{fresnel} \eea
In fact, it is a scalar equation for $k$ and thus implicitly
represents the dispersion relation for the light propagation in the
polarized QED vacuum. A suitable choice of gauge fixing for
$a_{\mu}$ simplifies the eigenvalue problem (\ref{eigen1}). We will
use a physical temporal gauge $a_0=\epsilon_0=0$, since it directly
links the polarization vector $\vec{\epsilon}$ to the electric field
of the propagating photon $\vec{e}$,
$\vec{\epsilon}=\vec{e}/|\vec{e}|$. With this gauge the eigenvalue
equation (\ref{eigen1}) splits into the equation
\begin{equation}
\label{E:00} A^{0i}\,\epsilon_i=0,
\end{equation}
and the reduced eigenvalue problem
\begin{equation}
\label{E:reduced} A^{ij}\,\epsilon_j=0\,.
\end{equation}
The latter implies the following condition
\begin{equation}
\label{E:detA=0} \mbox{det}\left(A^{ij}\right)=0.
\end{equation}
There are two independent physical modes of the eigenvalue problem
(\ref{E:reduced}), so that the space of polarizations is at most
two-dimensional \cite{birula,visser}.

\section{vacuum birefringence in Magnetic Field} \label{sec:magnetic}

In this section we first obtain the general equations for the light
velocity and polarization vector. Then we apply these equations to
both the truncated one-loop effective Euler-Heisenberg Lagrangian,
Eq.~(\ref{eq:EH2}), and the series representation for the one-loop
effective Lagrangian, Eq.~(\ref{L0}), in both the weak and strong
magnetic field regions. For the case of ultra strong magnetic field
we derive asymptotic formulae as well.

Without loss of generality, we choose the magnetic field directed
along the $z$-axis, $\vec{B} = (0, 0, a)$. We assume the wave vector
$\vec k$ lies in the  plane $xOz$, so that $ \bar k^\mu =k^\mu/ |
{\vec k} | = (v, \sin \theta, 0, \cos \theta)$ (Fig. 1), and we will
not distinguish between ${\bar k}$ and $k$ below. Hereafter the
coefficient functions $c_i$ in (\ref{deri}) are taken in the limit
of vanishing electric field, $b \rightarrow 0$ ($y \rightarrow 0$).
Since in standard QED the effective Lagrangian ${\cal
L}_{\mathrm{eff}}$ is an even function of $y$ we find
\bea c_4 = 0,~~~~c_5 =0. \label{c4c5} \eea

\begin{figure}[t]
\includegraphics[width=0.35\columnwidth]{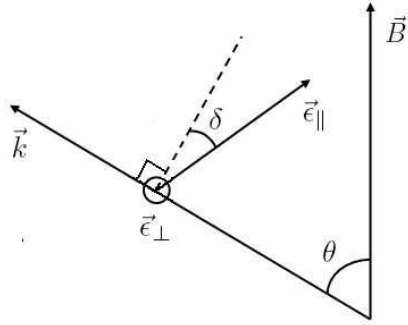}
\caption{\label{Fig. 0} Two modes of light propagation in a
magnetized vacuum. $\delta=\angle (\vec{\epsilon}_{\parallel},
\vec{k}) - \dfrac{\pi}{2}$. $\vec{\epsilon}_{\bot}$ is orthogonal to
the plane containing $\vec{B}$ and $\vec{k}$. }
\end{figure}

The explicit solution to the eigenvalue equation (\ref{E:reduced})
provides two independent polarization vectors,
$\vec{\epsilon}_{\bot}$ and $\vec{\epsilon}_{\parallel}$,
corresponding to the orthogonal and parallel modes respectively
\bea \vec{\epsilon}_{\bot}
&=& (0,~ 1,~ 0), \nn \\
v^2_{\bot} &=& 1+\frac{c_1 a^2 \sin^2 \theta}{c_3}, \nn \\
\vec{\epsilon}_{\parallel} &=& \dfrac{1}{\rho(\theta)} \big{(}(c_3
-a^2 c_2) \cos\theta, ~0, ~-c_3 \sin \theta \big{)}, \nn \\
v^2_{\parallel} &=& 1-\frac{c_2 a^2 \sin^2 \theta}{c_2 a^2 - c_3},
\label{formal} \eea
where $\rho(\theta)= \sqrt{c_3^2+a^2 c_2(a^2 c_2-2c_3)\cos^2
\theta}$ is the normalization factor.
One can check that the solution is consistent with Eq.~(\ref{E:00}).
It is worthwhile to notice that the polarization vector
$\vec{\epsilon}_{\parallel}$ is not orthogonal to the wave vector
$\vec{k}$. The deviation angle defined by $\delta=\angle
(\vec{\epsilon}_{\parallel}, \vec{k}) - \dfrac{\pi}{2}$ takes the
form
\bea \cot \delta = \cot \theta -\frac{2c_3}{a^2 c_2 \sin2 \theta}.
\eea
The existence of $\delta$ is analogous to the light propagation in
crystal optics, in which the non-orthogonality between the photon
electric field and wave vector often occurs. In some sense, the
vacuum in magnetic field behaves as a "uniaxial crystal".

Now we can apply the above formal equations to the one-loop
effective Euler-Heisenberg Lagrangian, Eq.~(\ref{eq:EH2}). Simple
calculation leads to the following results
 \bea
 v^2_{\bot} &=&
1-\frac{2a^2 \tilde {c}_1 \sin^2 \theta}{1-a^2 \tilde {c}_1}, \nn \\
v^2_{\parallel} &=& 1- \frac{2 a^2 \tilde {c}_2 \sin^2 \theta}{1-a^2(\tilde {c}_1-2\tilde {c}_2)}  ,\nn \\
\cot \delta &=& \cot \theta + \frac{1- a^2 \tilde {c}_1}{a^2 \tilde
{c}_2 \sin 2\theta}. \eea

We apply the above formal equations to exact one-loop effective
Lagrangian, Eq.~(\ref{L0}). One can calculate the coefficient
functions $c_i$ in terms of the main function $G(a)$ \cite{pak}
\bea c_1 &=& \dfrac{1}{2a^3}(a \pro^2_a {\cal L} - \pro_a {\cal L}), \nn \\
c_2 &=&  \dfrac{1}{2a^3}(a \pro^2_b {\cal L} + \pro_a {\cal L}), \nn \\
c_3 &=&  \dfrac{1}{2a} \pro_a {\cal L}, \nn \\
c_4 &=& \dfrac{1}{2a} \pro_b {\cal L}, \nn \\
\pro_a {\cal L} &=& -a -\dfrac{e^2 a}{2 \pi^4} G(a) - \dfrac{e^2 a^2}{4 \pi^4} G'(a) ,\nn \\
\pro_b {\cal L} &=& 0, \nn \\
\pro^2_a {\cal L} &=& -1 -\dfrac{e^2}{2 \pi^4} G(a) - \dfrac{e^2
a}{\pi^4} G'(a) -\dfrac{e^2 a^2}{
4 \pi^4} G''(a), \nn \\
\pro^2_b {\cal L} &=& 1 + \dfrac{e^4 a^2}{36 \pi^2 m^4} + \dfrac{e^4
a^3}{3 \pi^4 m^4} G'(a)+ \dfrac{e^4 a^4}{6 \pi^4 m^4}
G''(a),\label{deriv} \eea
where
\bea G(a) &=& \s \dfrac{1}{n^2} g(\dfrac{n \pi
m^2}{e a}), \nn \\
g(x) &=& {\rm ci}(x) \cos x + {\rm si}(x) \sin x  . \label{GGP} \eea
The function $G(a)$ determines the one-loop contribution to the
effective Lagrangian with a pure magnetic background \cite{pak}, and
it can be written in terms of the generalized Hurvitz
$\zeta$-function as well.

With the Eqs. (\ref{formal}) and (\ref{deriv}), the light velocities
and the angle $\delta$ can be expressed as follows
 \bea v^2_{\bot} &=&
1+\dfrac{\dfrac{e^2a}{4\pi^4} \sin^2 \theta (3G'(a)+a G''(a))}{1+
\dfrac{e^2}{4\pi^4}(2 G(a) + a G'(a))}, \nn \\
v^2_{\parallel} &=& {\Big(}1+ \dfrac{e^4 a^2 \cos^2 \theta}{36\pi^2
m^4}+
 \dfrac{e^2 \sin^2 \theta}{2\pi^4 }G(a)  \nn \\
 &&+(\dfrac{e^4 a^3 \cos^2 \theta}{3\pi^4 m^4} +
  \dfrac{e^2 a \sin^2\theta}{4\pi^4})G'(a) +\dfrac{e^4 a^4 }{6m^4 \pi^4}\cos^2 \theta G''(a) {\Big )}\nn \\
  && \cdot {\Big(}1+ \dfrac{e^4 a^2}{36\pi^2 m^4}+\dfrac{e^4 a^3}
  {3\pi^4 m^4}G'(a)  +\dfrac{e^4 a^4}{ 6m^4 \pi^4}G''(a){\Big)}^{-1}  ,\nn \\
\cot \delta&=&\cot \theta + \dfrac{ \csc \theta \sec \theta {\Big
(}1 +\dfrac{e^2}{2\pi^4}G(a) +  \dfrac{e^2 a}{4\pi^4}G'(a){\Big )}}
{ \dfrac{e^4 a^2}{36\pi^2 m^4}-\dfrac{e^2 }{2\pi^4 }G(a)+
(\dfrac{e^4 a^3}{3\pi^4 m^4}-\dfrac{ e^2 a}{4\pi^4}) G'(a)
+\dfrac{e^4 a^4}{6m^4 \pi^4}G''(a)}. \eea

\begin{figure}[t]
\includegraphics[width=0.8\columnwidth]{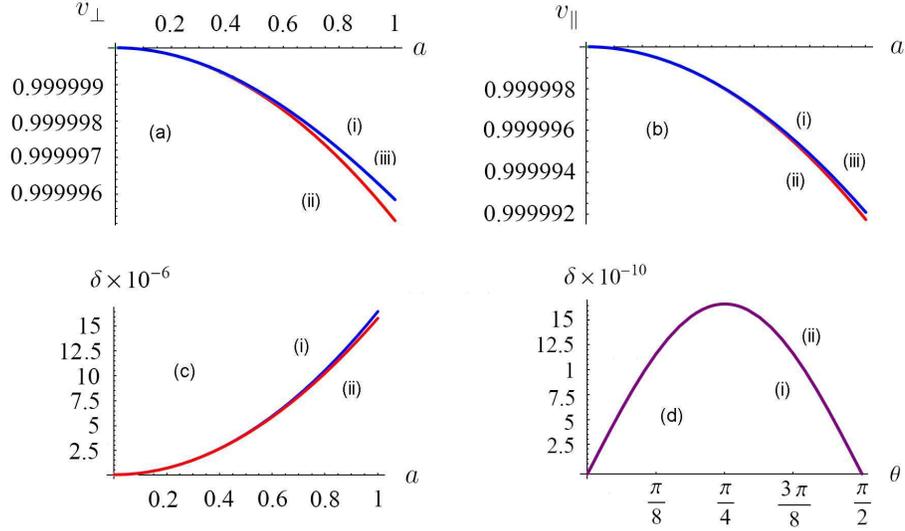}
\caption{\label{Fig. 1} Light propagation in weak magnetic field,
$\vec{B}=(0,0,a)$. (a) Light velocity of $\bot$ mode; (b) light
velocity of $\parallel$ mode; (c) the dependence of $\delta$ on the
field strength $\tilde a$ at $\theta=\dfrac{\pi}{4}$; (d) the
dependence of $\delta$ on $\theta$ at $\tilde a=0.01$ (the two
curves are quite adjacent). (i) an exact one-loop approximated
result; (ii) the result in weak field approximation; (iii) the
result obtained in Ref. \cite{pak}. The dimensionless magnetic field
strength $\tilde a=\dfrac{a}{m^2}$ is measured in units of electron
mass, the critical value is $\tilde a_c=\dfrac{1}{e}\simeq 3.3$. The
magnetic field strength $B$ in standard units is given by $B={\tilde
a} \times \sqrt{4 \pi \alpha} B_c $.}
\end{figure}

In order to confirm our results, we compare them with the results
obtained in the past for the particular case of vacuum birefringence
in weak field limit. For $\theta=\dfrac{\pi}{4}$ the light
velocities for ($\bot,
\parallel$)-modes are plotted in Fig. (2a) and (2b). The
dependence of $\delta$ on the field strength $a$ and on $\theta$ is
shown in Fig. (2c) and (2d) respectively.

For the case of weak field regime, the $\delta$ angle is quite small
as expected, $\delta \simeq a^2 \tilde {c}_2 \sin 2 \theta$. For
moderate magnetic fields satisfying the condition $-\dfrac{c_3}{c_2
a^2}\gg 1$ we have a simple relation
\bea \delta \simeq -\dfrac{c_2 a^2}{2c_3}\sin2\theta. \eea

Now, let us consider light velocity in the strong field region. From
the asymptotic behaviour of the function $G(a)$ \cite{pak}
\bea G(a) &=& -\dfrac{\pi^2}{6} ( \ln \dfrac {e a}{m^2} + d_1)
 -\dfrac{\pi^2 m^2}{2 e a} (\ln \dfrac{e a}{\pi m^2} +1)- \dfrac {\pi^2 m^4}{4 e^2 a^2} ( \ln \dfrac{2 e a}{\pi m^2}
-\gamma+\dfrac{5}{2}), \nn \\
d_1 &=& -\gamma - \ln \pi +\dfrac{6}{\pi^2} \zeta'(2)=- 2.29191...
~, \label{G1} \eea
one can derive the following asymptotic equations for the light
velocities $v_{\bot ,\parallel}$ and angle $\delta$ in an
ultra-strong magnetic field
\bea v^2_\bot &\simeq& \dfrac{ 1 - \dfrac{e^2}{12 \pi^2} ( \ln
\dfrac {e a}{m^2}
 + d_1+\dfrac{1}{2}+\sin^2 \theta) }{1 - \dfrac{e^2 }
 { 12 \pi^2} ( \ln \dfrac{e a}{m^2} + d_1+
\dfrac{1}{2})} \, \nn \\
&=& 1+{\cal O}(B_c/a), \nn \\
  v^2_\parallel
 &\simeq& \dfrac{ 1 - \dfrac{e^2}{12 \pi^2} ( \ln \dfrac {e a}{m^2}
 + d_1-d_2\cos^2 \theta+\dfrac{1}{2}) +\dfrac{ae^3}{12\pi^2m^2}\cos^2\theta }{ 1 -
  \dfrac{e^2}{12 \pi^2} (\ln \dfrac {e a}{m^2}
 + d_1-d_2+\dfrac{1}{2}) + \dfrac{e^3 a}{12 \pi^2 m^2}} \,\nn \\
 &=& \cos^2 \theta+{\cal O} (B_c/a),\nn \\
\cot \delta &\simeq& \cot \theta - \dfrac{ 2\ln \dfrac{a
e}{m^2}+1+2d_1 - \dfrac{24 \pi^2}{e^2}}
{\sin 2\theta(\dfrac{a e}{m^2}+d_2)}\,\nn \\
&=& \cot \theta-\dfrac{1}{\sin 2 \theta} {\cal
O}(B_c/a),\nn \\
  d_2 &=&
d_1-\dfrac{1}{2}+\gamma+\ln\dfrac{\pi}{2}=-1.76311...~.
\label{strong} \eea

\begin{figure}[t]
\includegraphics[width=0.8\columnwidth]{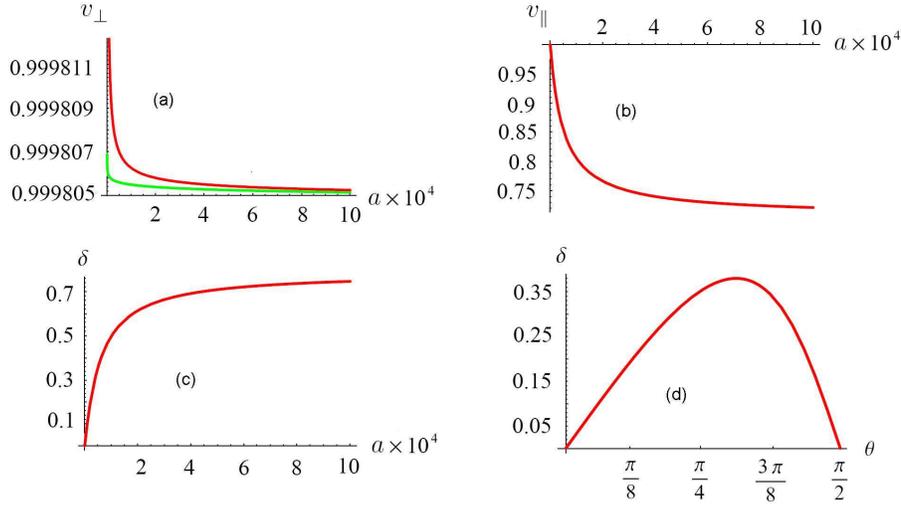}
\caption{\label{Fig. 2} Light propagation in strong magnetic field.
(a) Light velocity of $\bot$ mode, the upper is exact one-loop
approximated result while the lower is asymptotic result; (b) light
velocity of $\parallel$ mode; (c) the dependence of $\delta$ on the
field strength $\tilde a$ at $\theta=\dfrac{\pi}{4}$; (d) the dependence of
$\delta$ on $\theta$ at $\tilde a=5\times10^3$. (In (b), (c) and (d) the
two curves of exact result and asymptotic result are quite
adjacent).  The dimensionless magnetic field strength $\tilde a=\dfrac{a}{m^2}$ is
measured in units of electron mass.The maximal terms of numerically calculating
$G(a)$ is $2\times10^4$ for the maximal field.}
\end{figure}

Comparing these asymptotic formulae with the results obtained in
\cite{pak}, one can conclude that the velocities of orthogonal mode
coincide while the velocities of parallel mode differ essentially in
the asymptotic limit. The deviation angle $\delta$ can be quite
large within $[0,\dfrac{\pi}{2})$. When $\theta$ approaches the
value $\dfrac{\pi}{2}$ the angle $\delta$ vanishes, i.e., the light
becomes transverse.

It is well known that the strength of an electric field is limited
by the "Klein Catastrophe" while the strength of a magnetic field is
not. But there are several other physical limits that apply to
magnetic fields \cite{zaumen, lerche}. For example, diverse
interactions with photons and matter deplete energy and momentum
from the neutron star field, limiting its strength to
$B_{max}<10^{16} - 10^{18} G$ \cite{lerche}. This typical value of
order $10^{18}G$ determines the range of strength of considered
magnetic field.

In Fig. (3a) and (3b) the light velocities $v_{\bot,
\parallel}$) at $\theta=\dfrac{\pi}{4}$ are presented in the strong field regime. The
dependence of $\delta$ on the field strength $a$ and on the $\theta$
is shown in Fig. (3c) and (3d) respectively.

The results are still reasonable even in the asymptotic limit since
the phase velocities keep bounded, $0 \leq v_{\bot,\parallel} \leq
1$. In fact, the orthogonal mode propagates as in the trivial
vacuum, independent of the wave vector. On the other hand, the phase
velocity of the parallel mode is directly associated to the
direction of propagation. The propagation perpendicular to the
magnetic field ($v_{\parallel}(\theta=\dfrac{\pi}{2})=0$) is
strictly forbidden, while the parallel propagation is preferred
$\big{(}v_{\parallel}(\theta=0)=1 \big{)}$, in agreement with the
previous results \cite{dittbook}. As a result, photons in the
$\parallel$ mode eventually propagate along the magnetic field,
regardless of their incidence angle $\theta$. So that, since
$\delta=\theta$ for $\theta\in[0,\dfrac{\pi}{2})$ in the asymptotic
limit, the polarization vector of the $\parallel$ mode is mostly
directed along the $x$ axis (except for $\theta=\dfrac{\pi}{2}$),
irrelevant to the wave vector.

Using the equations for the light the velocities (\ref{formal}), one
can derive the corresponding refraction indices
\bea && n^2_{\bot , \parallel} = \dfrac{1}{v^2_{\bot ,
\parallel}} . \eea
With Eq.~(\ref{strong}) one can approximate the refraction index by
\bea n_\bot &\simeq& 1, \nn \\
 n_\parallel &\simeq& \big{(}
{\dfrac{1+\dfrac{e^3 a}{12 \pi^2 m^2}}{1+\dfrac{e^3 a}{12 \pi^2
m^2}\cos^2 \theta}}\big{)}^{\frac{1}{2}} , \eea
in agreement with results obtained before \cite{shabad,shabadnew}.

Our results can be helpful in study of the light propagation in a
pure electric field background or crossed field background
 ($\vec{E}\bot\vec{B}$,
$|\vec{E}|=|\vec{B}|$). For these two cases it is readily verified
that $y=0$ still holds and thus $c_4=c_5=0$. So the calculation is
straightforward by analogy with the above results. For example, in a
pure electric field, interchanging of $c_1$ and $c_2$ in
(\ref{formal}) will give the desired results.

\section{Conclusion}

We have analyzed the light propagation in a constant magnetic field
of arbitrary strength. Within effective action approach we have
investigated the features of the propagation modes in both the weak
and strong field regimes. We have demonstrated that the polarization
vector of the parallel mode is no longer orthogonal to the wave
vector. The effect of non-transversality is enhanced in strong field
regime and significantly affects the asymptotic behavior of light
velocity. The analytic asymptotic formulae of light velocities and
deviation angle $\delta$ for strong magnetic field have been
obtained.

We would like to discuss two potential applications of our results
in astrophysics. The first one is related to magnetic lensing effect
which appears when the fields are significantly stronger than $B_c$
(see, e.g., \cite{shaviv}). For example, for at least five known
gamma-pulsars the magnetic field exceeds $B_c$, and for magnetars
the magnitude of magnetic field is estimated to be of order
$10^{14}-10^{15}$G \cite{pulsar}. The main result of the lensing
effect is that the effective surface areas of the astrophysical
object measured by two polarization states are different. Since the
parallel mode is no longer transverse, the measurement of its
polarization responses accordingly. Especially, the dependence of
the deviation angle $\delta$ on the incidence angle $\theta$ should
be important in the determination of the effective surface area of
polarizations. This consequence in the measurement will be
strengthened in the strong magnetic field. From the numerical
results in Fig. 2, one may argue that the new feature of
non-transversality is negligible at $B\sim B_c$ compared to
traditional approach; however, for astronomical distances, even a
very small deviation can lead to essentially different observations.
Another possible application might be related to the effect of
strongly enhanced mode coupling in light scattering (see, e.g.,
\cite{bulik,ozel}). When photons propagate through scattering in the
magnetized plasma they can change their polarization modes as well
as their directions and energy. This effect can change the total
spectrum and angular distribution of radiation from the neutron
star. When the photons interact with both the electrons and the
protons in the plasma, a careful analysis of photon polarization
effects is necessary for precise calculation. We hope that our
results can provide a better quantitative description of these
effects and possibly other astrophysical phenomena related to
birefringence in ultra-strong magnetic fields.

  {\bf Acknowledgments}
  The authors would like to thank Prof. D. G. Pak for suggesting this
  work. We also thank Prof. Mo-Lin Ge and Prof. Yi Liao for their
  encouragement and helpful discussions.

\end{document}